\documentclass{article}

\usepackage{PRIMEarxiv}

\usepackage{float}
\usepackage[caption = false]{subfig}
\usepackage[export]{adjustbox}
\usepackage{caption}
\usepackage{listings,multicol}
\usepackage{lipsum}

\usepackage{todonotes}

\usepackage{amsmath,amsfonts}
\usepackage{algorithmic}
\usepackage{graphicx}
\usepackage{textcomp}
\usepackage{xcolor}

\definecolor{codegreen}{rgb}{0,0.6,0}
\definecolor{codegray}{rgb}{0.5,0.5,0.5}
\definecolor{codepurple}{rgb}{0.58,0,0.82}
\definecolor{backcolour}{rgb}{0.95,0.95,0.92}
\lstdefinestyle{mystyle}{
    backgroundcolor=\color{backcolour},   
    commentstyle=\color{codegreen},
    keywordstyle=\color{magenta},
    numberstyle=\tiny\color{codegray},
    stringstyle=\color{codepurple},
    basicstyle=\ttfamily\footnotesize,
    breakatwhitespace=false,         
    breaklines=true,                 
    captionpos=t,                    
    keepspaces=true,                 
    numbers=left,                    
    numbersep=4pt,                  
    showspaces=false,                
    showstringspaces=false,
    showtabs=false,                  
    tabsize=2
}

\definecolor{bgblue}{rgb}{0.41961,0.80784,0.80784}%
\definecolor{bgred}{rgb}{1,0.61569,0.61569}%
\definecolor{bggreen}{rgb}{0,0.91569,0.51569}%

\usepackage{mathtools}

\usepackage{soul} 
\title{SerIOS: Enhancing Hardware Security in \\Integrated Optoelectronic Systems
\thanks{\textit{\underline{Citation}}: 
\textbf{RSP 2023 - To be published}} 
}

\author{
  Felipe G\"{o}hring de Magalh\~{a}es, Gabriela Nicolescu \\
  Ecole Polytechnique de Montreal \\
  Montreal, QC \\
  Canada\\
  \texttt{\{felipe.gohring-de-magalhaes, gabriela.nicolescu\}@polymtl.ca} \\
   \And
  Mahdi Nikdast \\
  Colorado State University \\
  Fort Collins, CO \\
  USA\\
  \texttt{mahdi.nikdast@colostate.edu} \\
}

\begin{document}
\maketitle

\begin{abstract}
Silicon photonics (SiPh) has different applications, from enabling fast and high-bandwidth communication for high-performance computing systems to realizing energy-efficient optical computation for AI hardware accelerators. However, integrating SiPh with electronic sub-systems can introduce new security vulnerabilities that cannot be adequately addressed using existing hardware security solutions for electronic systems. This paper introduces SerIOS, the first framework aimed at enhancing hardware security in optoelectronic systems by leveraging the unique properties of optical lithography. SerIOS employs cryptographic keys generated based on imperfections in the optical lithography process and an online detection mechanism to detect attacks. Simulation and synthesis results demonstrate SerIOS's effectiveness in detecting and preventing attacks, with a small area footprint of less than 15\% and a 100\% detection rate across various attack scenarios and optoelectronic architectures, including photonic AI accelerators.
\end{abstract}


\section{Introduction}
    Silicon photonic (SiPh) integrated circuits and interconnects exploit the fast throughput of light-speed data transmission to realize ultra-high bandwidth communication with low-power dissipation in high-performance computing systems. In any optoelectronic system, such as an electronically programmable SiPh neural networks, the integration (i.e., monolithic, intra-chip, or inter-chip) of electronic and photonic sub-systems is necessary to deal with signals from different domains (i.e., optical and electrical). Such an integration also requires signal conversions and adaptations, as well as SiPh node reconfiguration for routing and computation~\cite{eu_ofc_2016}. 
   
    Systems integrating different technologies (e.g., an optoelectronic system) are more susceptible to malicious attacks where data can be stolen or system's normal behavior can be impacted~\cite{opt_mitigation3}. Attacks can be inserted by hardware Trojans (HT), which are malicious pieces of hardware designed to mischievously act and interfere with the targeted system. The impacts of attacks can include data leakage and manipulation, service and sleep-denial, irreversible losses, systemic abnormal behavior, and permanent breakdown~\cite{sergio}.  
    
    In optoelectronic systems with SiPh and CMOS electronic integration, attackers can take advantage of multiple signal conversions required to exchange data between the two domains and act on the integration interfaces. Furthermore, as the optical transmission behavior is directly related to SiPh device design characteristics (e.g., a waveguide width) and temperature, minimal changes in these characteristics can affect the transmission quality, where attackers have different ways to interfere with the underlying SiPh circuit. For example, considering the thermal sensitivity of SiPh devices, in the photonic accelerator proposed in \cite{clements}, an attacker (e.g., an IP integrated in the system) can increase its own temperature (e.g., by performing heavy mathematical operations). Consequently, this can impose thermal crosstalk~\cite{Kyatam:20}, generating noise or affect the neighboring SiPh devices (e.g., by imposing phase noise or frequency deviations) in proximity.  
    
    Existing techniques to address purely electronic hardware Trojan (e-HT) attacks might be explored for optoelectronic systems, but most will fail to address the hardware security concerns in optoelectronic systems~\cite{opt_mitigation}. While e-HTs target the IPs and electrical paths, optical HTs (o-HTs) target the optoelectronic interfaces and introduce disturbances to the SiPh sub-system. Prior efforts such as~\cite{opt_mitigation3} took a step further on presenting solutions to enhance the security of optoelectronic systems by taking into account the characteristics of the underlying optical path, adding an extra layer of security to the system design. However, these solutions are still limited because of their complex integration methods, limited applicability, and dependence on electronic sub-systems.  

    The novel contribution in this paper is on developing the first framework for enhancing hardware \underline{Se}cu\underline{r}ity in \underline{I}ntegrated \underline{O}ptoelectronic \underline{S}ystems, called SerIOS. SerIOS is designed to benefit from the unique characteristics of optical lithography processes to create unique seeds to generate unique keys. Furthermore, SerIOS interacts directly with the SiPh sub-system to enhance the signal transmission quality (e.g., by reducing noise). By dynamically analyzing transmitted optical power in the output, SerIOS is capable of detecting disturbances on an optical channel. Compared to similar prior efforts---such as~\cite{opt_mitigation3}, which benefits from physical unclonable functions (PUFs) and lithography process characteristics---SerIOS stands as the only solution that does not rely on error-prone off-line analyses, nor complex neural networks for pattern recognition.  Results, obtained from both simulation and FPGA prototyping, show that the online detection method has an area overhead of less than 15\% when the entire system's implementation (i.e., IPs, interface controllers, and SerIOS) is considered. Reported power consumption for target FPGA is as low as 200~mW and real-time executing latency is about 200~ns. More importantly, when randomly inserting o-HTs on a system for creating different attacks such as black-hole, sink-hole, and flooding, we show that SerIOS can achieve 100\% attack detection during runtime. 

\section{Background and Related Work}
\label{sec:back}
\subsection{Silicon Photonic Devices and Configuration}
\label{sec:base}
\begin{figure}[!t]
        \centering
            \includegraphics[scale=0.45,valign=c]{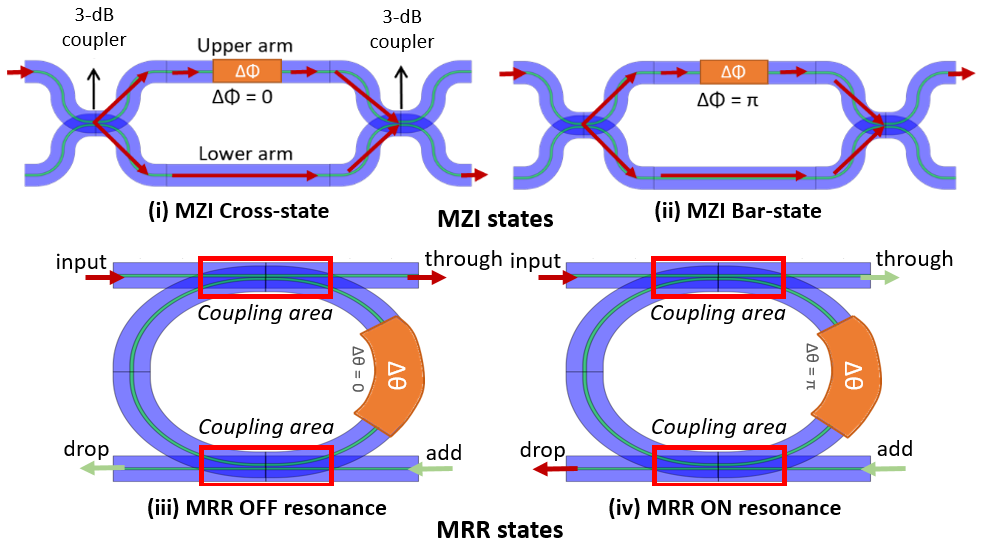}
            \caption{SiPh building blocks. MZI in two operation states: (i) Cross and (ii) Bar. MRR in two operation states: (iii) OFF resonance and (iv) ON resonance.}
        \label{fig:sip_nodes}
    \end{figure}

    Mach--Zehnder interferometers (MZIs) and microring resonators (MRRs) are commonly used building blocks in optoelectronic systems. Fig.~\ref{fig:sip_nodes} illustrates conventional MZI and MRR designs. MZIs are interferometric devices that consist of two 3-dB couplers and phase shifters on the arms~\cite{lukas}. By applying a phase shift, the optical signals passing through the MZI arms can experience destructive or constructive interference. For example, in the 2$\times$2 MZI shown in Fig.~\ref{fig:sip_nodes}, a phase difference of 0 results in constructive interference (Cross state), while a phase difference of $\pi$ leads to destructive interference (Bar state). Intermediate phase differences allow the signal to be split between output ports. MZIs find applications in optical switched networks and SiPh neural networks~\cite{clements}.
    MRRs are compact optical resonators formed by a closed loop waveguide, where light can be trapped and resonantly coupled. They can be used for filtering, modulation, and switching functionalities in optoelectronic systems. MRRs are particularly useful in wavelength division multiplexing (WDM) systems and integrated photonic circuits. These devices, MZIs and MRRs, play important roles in various optoelectronic applications, providing functionalities such as interference-based signal manipulation, wavelength filtering, and routing~\cite{dl_siph}.
    
    SiPh nodes operate based on their parameter configuration, which includes induced optical phase and tuned resonant wavelengths. These parameters can be dynamically and electronically adjusted to meet the requirements of the application. Usually, the controller node connects to each SiPh node and configures its operation accordingly. In the context of a coherent SiPh neural network \cite{mit_acc}, the underlying SiPh nodes (e.g., MZIs in \cite{mit_acc}) are adjusted in terms of optical phases by the electronic controller. These phase values represent the weight parameters in the neural network and can be determined using training and decomposition algorithms \cite{clements}. The electronic controller, designed for specific target applications, performs the necessary actions to achieve the desired system functionality~\cite{eu_ofc_2016}. It interacts with the SiPh sub-system through electro-optic interfaces, utilizing methods like thermal tuning (e.g., microheaters) or carrier injection (e.g., PN junctions). For instance, by applying a specific voltage to a microheater placed on top of an MZI arm (Fig.~\ref{fig:sip_nodes}, an electronic controller can configure the SiPh sub-system to alter the optical signal phase in the MZI.

    \subsection{Fabrication-Process Variations in Silicon Photonics}
    \label{sec:variations}
    Critical dimensions like width and thickness in SiPh devices, such as MZIs and MRRs, are affected by fabrication-process variations (PVs), leading to performance impact. PVs introduce deviations in power and phase output, compromising the behavior of SiPh sub-systems~\cite{9474000}. Systematic and random PVs pose challenges for accurate modeling, requiring runtime compensation solutions like active bias control \cite{4447311}. SiPh waveguide thickness, MRR response, and directional couplers are affected by PVs, causing malfunctions and performance degradation~\cite{9474000}. To mitigate PV impact, statistical modeling during the design phase or runtime approaches like dynamic bias control are employed~\cite{countering}. In SerIOS, we propose a real-time integrated method to detect and correct intentional deviations and attacks, reducing errors in optoelectronic systems.
    
\subsection{Physically Unclonable Functions (PUFs)}
\label{sec:puf}
    A physical unclonable function is a component that utilizes the inherent randomness introduced during circuit manufacturing to generate a unique fingerprint for a physical entity \cite{puf}. In the case of fabricated SiPh circuits, variations in their parameters and performance, as discussed in Section~\ref{sec:variations}, result from unpredictable factors such as random physical deviations. When designing SiPh-based PUFs, there are different approaches to consider. Some works, such as \cite{9465434}, focus on designing SiPh sub-systems specifically engineered to produce PUFs. Another approach is to utilize the impact of PVs on the SiPh sub-system, such as optical interconnects and accelerators, and extract the unique characteristics of transmitted signals (e.g., optical intensity) to create PUFs. SerIOS adopts the latter approach, avoiding the need to modify the SiPh sub-system with specific PUF modules.

\subsection{\small Security Breaches and Threat Models in Optoelectronic Systems}
    Threats to optoelectronic systems can arise from both software and hardware components. The hardware block can be compromised by a Hardware Trojan (HT) designed to maliciously impact the system. These HTs can be inserted by a malicious engineer during the design phase or in compromised foundries by modifying the design layouts. Software attacks can originate from infected IPs integrated with the SiPh sub-system. HT attacks on an optoelectronic system typically involve four phases: Design Phase, Activation Phase, Operation Phase, and Idle Phase \cite{25930692593144}. 

  \begin{figure}[!t]
        \centering
            \includegraphics[scale=0.6,valign=c]{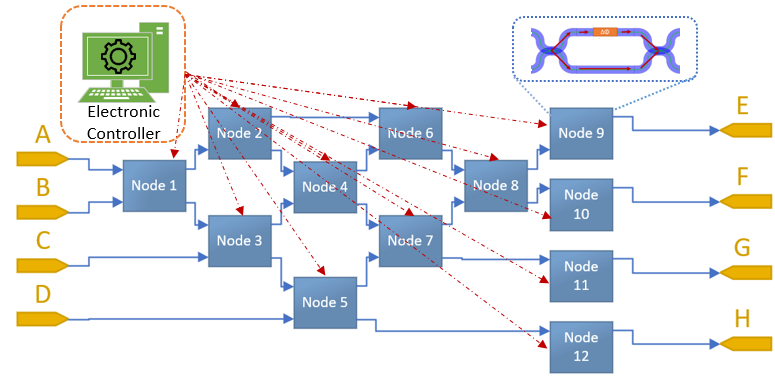}
            \caption{An optical linear multiplier based on MZI nodes to perform optical-domain matrix-vector multiplication~\cite{mit_acc}.}
        \label{fig:mit_overview}
    \end{figure}

    We consider HTs introduced in the foundry, affecting the entire system. Malicious applications at the IP level can also introduce various types of attacks. Side-channel attacks exploit one-way interference to cause information leakage by analyzing the impact of interference on the system. As an example architecture, we consider the SiPh-based optical multiplier depicted in Fig.~\ref{fig:mit_overview}, based on the accelerator architecture proposed in \cite{mit_acc}. Each SiPh node is based on MZI, similar to the one shown in Fig.~\ref{fig:sip_nodes}. The architecture includes connections between nodes, I/Os, and an electronic controller connected to each MZI node. In this system, attackers can launch attacks from different perspectives. For example, a neighboring IP integrated with the SiPh sub-system can manipulate the signal phase values by increasing its temperature and affecting the nearby SiPh MZIs. In the presence of a HT, the propagation direction of the optical signals can be manipulated, thereby altering the signal phase values. These attacks directly impact the output readings, leading to interference with the application behavior and potentially degrading system performance. Such deviations can lower the system's inferencing accuracy \cite{onn_under}.

    To mitigate these attacks, various solutions can be employed, such as fingerprinting and active detection mechanisms. In SerIOS, we adopt an online detection strategy where a baseline system operation is established. Periodic verification is performed on the SiPh sub-system to ensure its security and correctness. 
    
\section{S\lowercase{er}IOS Security Framework}
\label{sec:serios}
\underline{Se}cu\underline{r}ity in \underline{I}ntegrated \underline{O}ptoelectronic \underline{S}ystems (SerIOS) framework, enables an online detection of anomalies on the SiPh sub-system and provides a method for unique cryptographic key generation and proper tuning of the underlying SiPh devices. It operates as a standalone hardware unit, enabling agnostic surveillance of the SiPh sub-system. Due to its modular design, SerIOS can be easily integrated with FPGA designs as a security layer.

\begin{figure}[!t]
\centering
    \includegraphics[scale=0.5,valign=c]{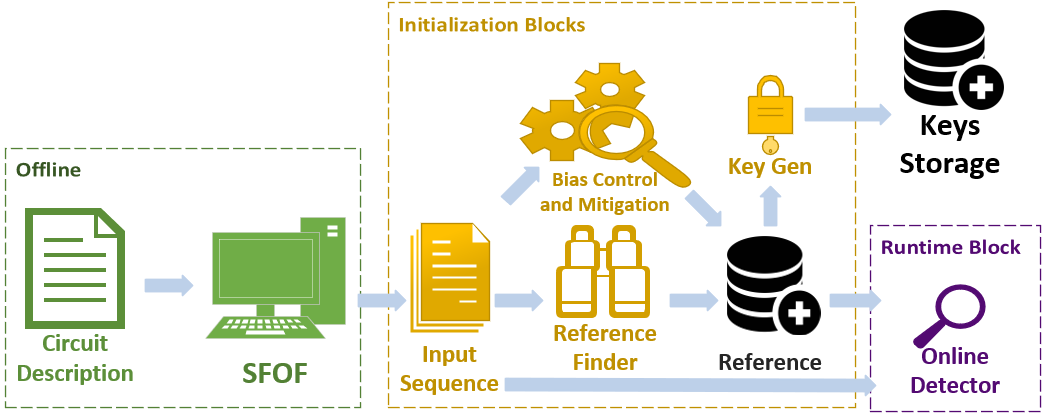}
    \caption{An overview of different building blocks in SerIOS.}
\label{fig:overview}
\end{figure}

Fig.~\ref{fig:overview} shows a global overview of the different building blocks and integrated modules in SerIOS. SerIOS execution is divided between offline, initialization blocks, and runtime blocks. Initialization and runtime blocks are the hardware blocks that constitute SerIOS, while offline execution is performed during design time and is not part of the integrated hardware blocks. Initialization blocks are executed once during system initialization and then are put in idle, while runtime blocks are  responsible for attack detection. The rest of this section describes the different building blocks of SerIOS.

\subsection{Straightforward Order Finder}
\label{sec:sfof}
The Straightforward Order Finder (\textit{SFOF}) analyzes the SiPh sub-system, looking for the correct tuning order of SiPh devices and nodes. This order is important because, as the optical signal travels through the SiPh sub-system, the effects of PVs will be accumulated through each subsequent SiPh node on the path. For instance, considering Fig.~\ref{fig:mit_overview} and input A, SiPh node 1 deviations will affect not only node 1, but also subsequent SiPh nodes to its right (e.g., if we consider MZIs in Bar state in all the nodes, the following nodes will be affected: 1, 2, 4, 6, 8, and 9). This way, \textit{SFOF} works based on a high-level description of the circuit and simulates the possible routes and paths to establish the optimized detection order.

\textit{SFOF} performs two main tasks: sequence generation and finding baseline communication patterns. For the \textbf{sequence generation}, \textit{SFOF} analyzes the SiPh sub-system topology based on the high-level circuit description. It determines the sequences that the Bias control and mitigation module should follow to find the readjusting tuning parameters of the SiPh nodes during runtime. To find the sequence, SFOF simulates the SiPh sub-system offline by emulating SiPh node transmissions while changing the tuning/configuration parameters of each node. The simulation is performed in reverse order, from the rightmost to the leftmost SiPh nodes, to capture the transmission path of each node based on the output triggered by that node. The generated sequence includes tuples of "node-output(s)" associations, capturing the relationships between nodes and the affected outputs.

For the \textbf{Baseline Communication Patterns}, SFOF identifies the communication patterns used later by the Reference Finder. It aims to find a set of communication pairs where each input is associated with at least one output. These patterns reflect the passive behavior of the SiPh sub-system when nodes are not tuned. For example, in the circuit of Fig.~\ref{fig:mit_overview}, assuming each SiPh node connects its inputs straight to its outputs (nodes in Bar state), the identified communication patterns are input A triggering output E, input B triggering output G, input C triggering output F, and input D triggering output H. These patterns represent the expected behavior of the SiPh sub-system and serve as a reference for detecting anomalies.

The goal of SFOF is to reduce runtime complexity and logic hardware overhead. By performing offline analysis and storing relevant information in a secure memory, SFOF facilitates its integration with the target system while minimizing runtime processing. However, these offline steps are not mandatory during design time and can be integrated into SerIOS initialization blocks for analysis during the initialization process, albeit resulting in longer initialization time.

\begin{lstlisting}[belowskip=-0.9 \baselineskip, language=Python, caption=SerIOS \textit{BCM} execution flow., label=lst:node_tuning, float=b]
tuningValue = [0]*numberOfNodes
for each node i in SiP_Circuit:
    outputPort = affectedPort(node_i)
    referenceMetric = nodeMetric(outputPort)
    tuningValue[node_i] += 1
    while referenceMetric < nodeMetric(outputPort):
        tuningValue[node_i] += 1
    saveValue(node_i, tuningValue[node_i])
\end{lstlisting}

\subsection{Bias Control and Mitigation Unit}
\label{sec:tuning}
In SerIOS, the \textit{Bias Control and Mitigation (BCM)} module is utilized to determine the appropriate configuration parameters for each SiPh node, taking into account the impact of PVs. The BCM module continuously monitors the operation of nodes within the SiPh sub-system and dynamically adjusts the node tuning parameters to compensate for the effects of PVs during runtime. While the BCM module can also address thermal variations, this work focuses specifically on PVs. The configuration parameters chosen by the designer, such as current/voltage requirements or phase shifts, are individually adjusted for each SiPh node. Real-time readings from the output ports, specifically the photodetector readings, are used to identify the optimal configuration values that mitigate the influence of PVs. It is important to note that SerIOS does not require readings from every individual node but relies on output port readings to achieve its objectives.   

The \textit{BCM} module in SerIOS is fed by the \textit{Input Sequence} generated by the \textit{SFOF} block. It collects relevant information and incorporates it into the online blocks to optimize their operation and reduce overhead. Pseudo-code for the \textit{BCM} module is provided in Listing \ref{lst:node_tuning}. The module iterates through each node, noting the affected port(s) based on the \textit{SFOF} sequencer, and the desired metric (e.g., phase shift) on those port(s). Next, the module incrementally increases the metric under evaluation by its minimal step (e.g., incrementing the temperature by 0.1\textsuperscript{o} C for a microheater) and assesses the resulting output of the SiPh sub-system. This process is repeated until a defined condition is met, which could be a predetermined optical power requirement on an output port. The tuning value obtained during this procedure is then stored in the \textit{References} database. This initialization process considers the conditions, such as temperature, at the time of execution. If these conditions change, the same procedure can be repeated to update the tuning values accordingly. 

\subsection{Unique Key Generation}
\label{sec:key}
The module responsible for key generation takes as input the values annotated by the \textit{BCM} module and utilizes mixing functions to create unique keys. This is possible because the PV-induced deviations in the SiPh sub-system are unique, resulting in unique tuning values (as discussed in Section \ref{sec:variations}). The validation functions used are designed for simple \textit{hardware} implementation, requiring minimal resources and relying on shift and logical operations only.

The \textbf{key uniqueness} is ensured by the nature of PVs in SiPh integrated circuits, as extensively analyzed and quantified in prior work~\cite{mahdi_pv}. PVs introduce unpredictable changes to the circuit, leading to unpredictable system behaviors. As discussed in Section \ref{sec:tuning}, the impact of PVs can be directly observed in the transmitted signals, which can be measured during runtime and are unique to each SiPh sub-system. The keys are generated based on these unique and unpredictable changes. Since these variations differ from one SiPh node to another (even within the same device, such as the two arms of an MZI), each circuit composed of a series of these nodes will have a unique set of variations and affected behavior. As demonstrated in previous work \cite{puf}, leveraging these unique PVs for each circuit allows the generation of unique seeds, thereby justifying the reliability of the keys.

For the \textbf{key generation} process, shift and logical operators are used to simplify the hardware implementation. These functions can be easily modified and function as standalone blocks. An example of key generation function is $$h(x) \Leftarrow \forall b_i, b_i\neq~b_{(i+1)}, XOR(b_i, b_{i-1}) \ll OR(b_i,b_{i+1}),$$ where each bit of a tuning parameter is evaluated, and their concatenation is performed using defined logical functions. Despite its simplicity, this function can generate unique keys thanks to the unique input seed used.

In SerIOS, different keys are generated using different functions and stored in the \textit{Secure Storage}. The \textit{Secure Storage} can be an isolated memory accessible only by SerIOS or a memory with additional security layers~\cite{34478183460374}. SerIOS can generate keys dynamically during runtime, considering the low latency overhead. The \textit{Key Gen} module creates keys for each communicating pair, allowing IPs to encrypt their transmissions. The number of keys generated and their usage can vary based on application requirements. SerIOS can generate keys based on implemented functions and different input seeds. Simple key generation functions are used to minimize hardware and execution overhead. In the studies conducted, one key function per transmitting pair was used, generating keys once and storing them. However, key generation can be done differently, such as generating keys each time they are used or using the same function for multiple scenarios. The security relies on the uniqueness of the seed. Implementing different functions has negligible impact on hardware and execution time. The key generation functions are designed to be simple, using basic operators, and their footprint is small compared to the entire SerIOS circuitry. This is why a function per transmitting pair was chosen, providing flexibility without significant overhead.

\begin{lstlisting}[belowskip=-0.9 \baselineskip, language=Python, caption=SerIOS online detection execution flow., label=lst:online, float=b]
for each pattern i in referenceValues:
    outputPort = affectedPort(pattern_i)
    reference = referenceValues(pattern_i)
    config = configuration(pattern_i)
    configureCircuit(config)
    reading = readMetric(outputPort)
    if difference(reference, reading) > threshold:
        reportIssue(pattern_i)
\end{lstlisting}

\subsection{Golden Values}
\label{sec:golden}
SerIOS's detection mechanism utilizes baseline execution scenarios, known as \textit{golden values}, which are defined during the design phase. These scenarios provide predetermined operational conditions, inputs, outputs, and configuration parameters. During system initialization, SerIOS captures these golden values by executing the predefined scenarios generated by the SFOF block, as discussed in Section~\ref{sec:sfof}. By configuring the SiPh sub-system accordingly, SerIOS obtains readings of optical signal powers and phases. Executing this step during system initialization ensures the isolation of the SiPh sub-system from other IPs, minimizing potential interference and preserving the integrity of the readings. While it is possible to repeat this step if the SiPh sub-system undergoes rearrangement, such as node failure or deactivation, re-initializing the sub-system with the same parameters is not explored in this paper. The focus of this work is on obtaining the initial golden values for future reference, rather than addressing specific scenarios of reconfiguration.

\subsection{Online Detection Methodology}
\label{sec:detection}
The online detection mechanism in SerIOS can be configured to operate continuously in parallel with ongoing transmissions or at predefined intervals. In this paper, the online detection is executed at specific intervals. To ensure compatibility with regular system operation, the same transmission wavelength is used for both data transmission and checking. This allows the system to verify different scenarios within the same wavelength band. The pseudo-code in Listing \ref{lst:online} outlines the methodology of the online detection module. It compares runtime readings with the reference values stored in the \textit{reference} database. The module iterates through each stored scenario, replicating the same inputs and following the same SiPh sub-system path. The output port used for comparison is also the same. If the difference between the new reading and the corresponding golden value exceeds a predefined threshold, SerIOS triggers an alarm to indicate an unusual situation. The threshold can be defined based on application-specific metrics, such as precision reduction in a multiplier circuit or the signal-to-noise ratio (SNR) at the SiPh sub-system output in a communication context.

To ensure the validity of readings and comparisons, the threshold should not be affected by external factors. For example, in a multiplier circuit, the MZI phases and inputs are used to establish the threshold, isolating it from potential external influences. However, it should be noted that thermal variations in the SiPh circuit can still result in false positives, impacting system operation. One possible solution to detect and adapt to temperature variations is to employ MRR-based thermal sensing, as described in \cite{opt_mitigation}. This technique estimates temperature variance and considers its impact on the readings. However, mitigating thermal variations is not the primary focus of this work. Additionally, white noise attacks can interfere with the accuracy of the results. Nevertheless, SerIOS, being aware of the nature of such attacks and the original state of the circuit, can identify them as threats and report them accordingly.

\begin{figure}[!t]
        \centering
        \includegraphics[scale=0.75]{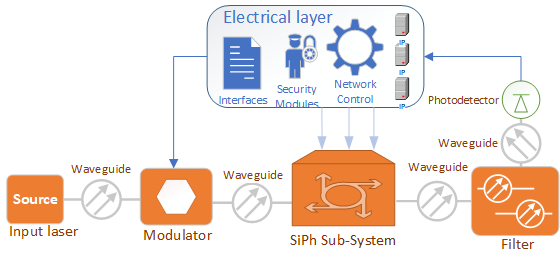}
        \caption{Validation environment presenting the main blocks composing the testbed.}
        \label{fig:test_architecture}
    \end{figure} 
    
\section{Results and Discussions}
\label{sec:results}
In this section, we consider different case studies to verify various aspects of SerIOS, based on different attacks that impact the system differently: \textbf{Black-hole:} The SiPh node is configured in such a way that all traffic is dropped; \textbf{Sink-hole}: All transmission is directed to a target node, saturating the transmission; \textbf{Flooding:} Non-stop inputs flood the optical channels, compromising the transmission; \textbf{Rerouting:} The SiPh nodes' tuning parameters are altered to route the transmission differently; and, \textbf{IP Hick-jacking:} An IP integrated to the SiPh sub-system maliciously interferes with it (i.e., increasing its own temperature). Lumerical Interconnect~\cite{inter} is used for modeling and simulation of the case studies. The evaluation scenarios are designed to simulate the validation platform presented in Fig. \ref{fig:test_architecture}. To insert the variations on SiPh nodes, we have used  canonical random generators to emulate the impacts of PVs for each node. 

\subsection{SerIOS Validation}
\begin{figure}[!t]
    \centering
        \includegraphics[scale=0.55,valign=t]{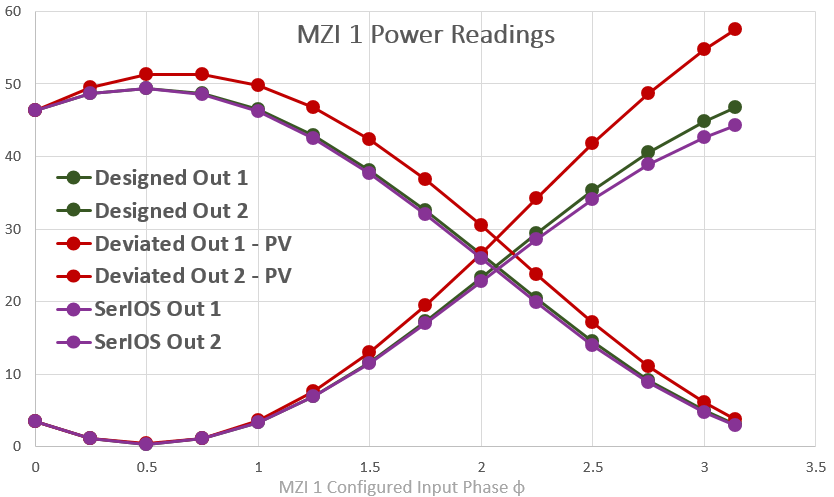}
        \caption{Optical output readings for one MZI (Nodes 1, considered as an example) composing the architecture in Fig. \ref{fig:mit_overview}.}
    \label{fig:mzi1-12}
\end{figure}

To illustrate SerIOS flow and performance, we use the SiPh sub-system presented in Fig. \ref{fig:mit_overview} for discussion. SiPh nodes are manipulated---e.g., MZI nodes are modified such that the phase shifter operates wrongly, emulating variations---and the affected output signals are measured. The SiPh sub-system in Fig.~\ref{fig:mit_overview} is composed of 12 identical MZIs, similar to the one presented in Fig.~\ref{fig:sip_nodes}. To control each MZI, a phase shifter is used. Fig~\ref{fig:mzi1-12} presents the optical power readings for one out of the 12 MZIs in the SiPh sub-system in Fig.~\ref{fig:mit_overview}, chosen arbitrarily to illustrate SerIOS impact on the node tuning. As it can be seen, SerIOS' tuned outputs are much closer to the ideal ones, when compared with deviated outputs. 

Bias control and mitigation and initial analysis are executed once during system initialization, and hence the latency of these do not impact execution time. The execution time to perform these initial steps is directly related to the evaluation of configuration parameters and the number of SiPh nodes that has to be analyzed. Another aspect affecting the latency is the delay for opto-electrical interface conversions. Furthermore, the delay for SiPh node stabilization after tuning should be considered. Accordingly, the overall latency can be modeled as $$t_{tun}(i) = \sum_{0}^{n} \rho.(\upsilon+\varsigma),$$ in which $\rho$ is the number of executions needed for the used configuration parameter, $\upsilon$ is the delay for SiPh node stabilization, $\varsigma$ is the opto-electrical conversion delay, and \textit{n} is the total number of reconfigurable SiPh nodes in the system. Taking as an example the circuit in Fig.~\ref{fig:mit_overview}, if the tuned parameter is the phase change and the phases window varies from $-\pi$ to $\pi$ in intervals of 0.1 rad, 629 different values should be verified for each one of the MZI nodes (7548 in total). Assuming the MZI switch time of 6~ns~\cite{eu_ofc_2016} and the opto-electrical conversion delay of 4~ns (one clock cycle in our validation environment), the overall latency to find the SiPh sub-system tuning parameters is 75.48~$\mu$s. Note that the module execution time can be ignored as it can execute in parallel with the signal converters and during stabilization time.

Next, golden values are collected. In this case, four transmission scenarios (i.e., four different phase configurations for the MZIs) are used as the configuration baseline. These scenarios are randomly chosen and each configuration parameter is arbitrarily chosen. For the creation of unique seeds, values collected by the \textit{BCM} unit are used. In this validation scenario, we have opted to create four different seeds, one associated with each input. This way, four seeds and four key-creation equations are used. Generated keys are internally stored within SerIOS and will be used later for encrypting the transmitted signals. For evaluating SerIOS performance and detection mechanism, we have manually inserted different types of attacks in the system, as presented in the introduction of this section. We insert each of these attacks at an arbitrary simulation time and verify SerIOS' responses. Table \ref{tab:attacks_detection} illustrates the output readings for each scenario, as well as the expected readings. As it can be seen, each attack interferes in a different manner with the SiPh sub-system. With this information, SerIOS is able to detect the anomaly and detect a possible attack situation. Encrypting messages with they generated keys, SerIOS protected transmissions against eavesdropping and spoofing attacks. PRBS sequences were used to validate it, where even for situations when attackers could retrieve the transmitted information, it was encrypted and hence useless to the attacker.

\begin{table}[!t]
    \footnotesize
           \centering
           \caption{Impacts on the output readings for SerIOS attacks detection. Each attack interferes with different output readings. The number in parentheses show the targeted SiPh node.}
            \label{tab:attacks_detection}
            \begin{tabular}{ccccc}
                                    & \multicolumn{4}{c}{\textbf{Outputs (dBm)}}                        \\ \cline{2-5} 
                                    & \textbf{E}     & \textbf{F}     & \textbf{G}     & \textbf{H}     \\ \hline
            \textbf{Golden Value}   & \textit{29.46} & \textit{30.57} & \textit{28.60} & \textit{34.60} \\ \hline
            \textbf{Black-hole (4)} & \textit{26.67} & \textit{31.84} & \textit{31.02} & \textit{34.60} \\ \hline
            \textbf{Sink-hole (5)}  & \textit{25.54} & \textit{27.73} & \textit{23.36} & \textit{39.12} \\ \hline
            \textbf{Flooding (1)}   & \textit{29.19} & \textit{31.60} & \textit{28.59} & \textit{34.03} \\ \hline
            \textbf{Rerouting (9)}  & \textit{0.0}   & \textit{30.57} & \textit{28.60} & \textit{34.60} \\ \hline
            \textbf{IP Hichjacking} & \textit{30.17} & \textit{30.68} & \textit{26.87} & \textit{34.70}
            \end{tabular}
\end{table}

\subsection{SerIOS Detection Efficiency and Overhead}
We evaluate SerIOS using different photonic switch architectures~\cite{benes}\cite{spanke} ranging in size from 2$\times$2 to 12$\times$12, as well as state-of-the-art optical neural network architectures~\cite{mit_acc}\cite{clements}\cite{mcgill_acc}. To generate the scenarios, we use canonical random generators to introduce variations in the system and simulate the impact of attacks. We focus on the direct consequences and outcomes of the attacks on the SiPh sub-system rather than the underlying causes. Each scenario is simulated 100 times to ensure accuracy and stability in the performance of SerIOS modules. 

\subsubsection{HT Attacks and Detection}
For evaluating the runtime detection efficiency of SerIOS, five attack scenarios are created using a random seed. These attacks simulate the impact of malicious nodes on the circuit. The SiPh sub-systems are modified to include the following attacks: black-hole, sink-hole, flooding, IP hijacking, and rerouting. Rather than having continuous attacks, each attack is randomly triggered to affect the SiPh sub-system. This creates asynchronous behaviors, emulating attackers that act under certain conditions. To achieve this, an execution window is defined (e.g., 10 seconds of system execution), and attacks are randomly triggered within this window. This means that an attack can occur at any time, multiple times during each window.

During both automated and manual tests, SerIOS successfully detects 100\% of the anomalies. False positives can happen, and acknowledging that it could lead to unwanted alarms, a solution to detect and adapt to the temperature variation is to use MRR-based thermal sensing~\cite{opt_mitigation}. It's important to note that SerIOS does not precisely locate the exact SiPh node under attack. The goal of SerIOS is to be agnostic to access and configuration controls of the SiPh sub-system, such as a network routing controller for a SiPh network sub-system. However, by integrating a network routing controller and configuring different transmission setups, it becomes possible to isolate and identify the specific node causing the issue by transmitting data while avoiding that node.

\subsubsection{Overhead Analysis}
As most of the processing in SerIOS is performed offline by \textit{SFOF}, the integrated modules remain simple circuits and can be scaled up for large-scale SiPh sub-systems (i.e., larger number of integrated nodes). The information for the golden values can be stored in a few registers and the remaining information locally within the integrated circuit. To evaluate the area overhead, two different scenarios are considered: first, SerIOS integration with an electronic controller and as part of an integrated system (i.e., a system-on-chip (SoC)); second, SerIOS scalability considering a higher number of I/Os and SiPh nodes in the circuit. For this, \cite{benes,spanke} switch architectures with sizes varying from 2$\times$2 to 12$\times$12, and optical neural network architectures presented in \cite{mit_acc,clements,mcgill_acc} are considered. SerIOS is synthesized for Xilinx Kintex 7 FPGA with a target 4~ns clock period for all the designs. Optimization was set to balance between area and speed. 

\begin{table}[!t]
    \centering
    \caption{SerIOS SoC integration footprint analysis. FPGA LUT and Flip-flop usage is considered for SerIOS, an electronic controller~\cite{hyco}, and a processing core~\cite{hfrisc}.}
            \label{tab:fpga}
            \begin{tabular}{lcccc}
            \multicolumn{5}{c}{\textbf{Integrated System Area Overhead}}              \\ \hline
             & \textbf{SerIOS} & \textbf{Others} & \textbf{Total} & \textbf{\%} \\  
            \textbf{SerIOS + \cite{hyco}} & \textit{1811} & \textit{1695} & \textit{3506}& \textit{51 \%} \\
            \textbf{SerIOS + \cite{hyco} + \cite{hfrisc}} & \textit{1811} & \textit{9587} & \textit{11398} & \textit{15 \%} \\ 
            \end{tabular}
\end{table}

Table~\ref{tab:fpga} shows SerIOS SoC integration with an electronic controller~\cite{hyco} and low-area processing IPs~\cite{hfrisc}. As illustrated, the footprint to integrate SerIOS accounts for less than 15\% of the entire system. Considering SerIOS scalability, Table~\ref{tab:fpga_serios} presents the synthesis results for different SerIOS configurations as well as the static and dynamic power consumption. Different SiPh switch fabric circuits are used where the overhead increases with the number of secured nodes as well as with the number of I/Os. This is due to the fact that SerIOS stores the golden values for each I/O transmission pair which grows with the number of I/Os. 

To evaluate execution time overhead, three modules should be considered: the \textit{BCM}, the \textit{online detection}, and the \textit{key generation}. The \textit{BCM} latency is observed only during system initialization, as this module is not to be executed continuously. Nevertheless, if it is to re-execute for a given reason (i.e., system re-calibration), $$t_{tun}(i) = \sum_{0}^{n} \rho.(\upsilon+\varsigma)$$ can be used to compute the execution time. Concerning the \textit{key generation}, in our studies, thanks to the fact that simple operators are used, the \textit{key generation} takes one clock cycle per function (i.e., in our environment, each clock cycle is 4~ns). For the \textit{online detection}, its execution overhead is related to the number of I/Os such as that the same communication combinations found by \textit{SFOF} (see Section \ref{sec:sfof}) are used. Accordingly, SerIOS latency can be defined as $$t_{det} = \sum_{0}^{n}\iota+\upsilon+\varsigma,$$ where $\upsilon$ is the SiPh node stabilization time, $\varsigma$ is the opto-electrical conversion time, \textit{n} is the number of communication combinations, and $\iota$ is the time overhead to access the golden value storage (i.e., \textit{Reference} database). Taking as an example the circuit in Fig.~\ref{fig:mit_overview}, assuming the MZI stabilization delay is 6~ns and the opto-electrical conversion takes 4~ns, and considering four communication combinations and the delay to access the golden values to be two clock cycles (i.e., using our environment, each clock cycle is 4~ns), the total time to check the system is 228~ns (i.e., 57 clock cycles).

\section{Conclusion} 
\label{sec:conclusion}
   Silicon photonics (SiPh) is emerging to boost the communication and computation performance in high-performance computing systems. Nevertheless, SiPh and CMOS electronic integration opens new doors for attackers, making the optoelectronic systems prone to security attacks. This work presented a novel framework comprising security breaches detection allied with mitigation techniques for process variations in optoelectronic systems. The proposed \underline{Se}cu\underline{r}ity in \underline{I}ntegrated \underline{O}ptoelectronic \underline{S}ystems (SerIOS) framework enables real-time detection and unique key generation for encryption algorithms. Results showed that SerIOS is able to dynamically evaluate SiPh nodes and properly set the tuning parameters for each of them. Furthermore, SerIOS real-time detection demonstrated 100\% detection of test cases during execution time. This work shows the importance of hardware security in emerging optoelectronic systems and highlights the need for interdisciplinary research efforts to address security concerns in such systems.   

\begin{table}[!t]
    \centering
    \caption{SerIOS scalability analysis. FPGA LUT and Flip-flop (FF) usage is considered for SerIOS as it scales with the SiPh sub-system.}
    \label{tab:fpga_serios}
    \begin{tabular}{lccccc}
    \multicolumn{6}{c}{\textbf{SerIOS Scalability Analysis}}              \\ \hline
    & \textbf{I/Os} & \textbf{Nodes} & \textbf{LUTs} & \textbf{FFs}  & \textbf{Power (mW)}\\  
    \textbf{SiPh SW \cite{spanke}} & \textit{4} & \textit{5} & \textit{843} & \textit{972} & \textit{215} \\
    \textbf{SiPh SW \cite{benes}} & \textit{4} & \textit{6} & \textit{822}& \textit{989} & \textit{216} \\
    \textbf{SiPh ACC \cite{mit_acc}} & \textit{4} & \textit{12} & \textit{1107}& \textit{1158} & \textit{202} \\
    \textbf{SiPh ACC \cite{mcgill_acc}} & \textit{4} & \textit{10} & \textit{1081}& \textit{1124} & \textit{201} \\
    \textbf{SiPh SW \cite{benes}} & \textit{6} & \textit{12} & \textit{1795}& \textit{2239} & \textit{247} \\
    \textbf{SiPh ACC \cite{clements}} & \textit{9} & \textit{36} & \textit{4814}& \textit{7069} & \textit{442} \\
    \textbf{SiPh SW \cite{benes}} & \textit{12} & \textit{28} & \textit{7100}& \textit{10817} & \textit{610} \\
    \end{tabular}
\end{table}
   
\bibliographystyle{unsrt}
\bibliography{bibliography}

\begin{thebibliography}{10}

\bibitem{eu_ofc_2016}
{Y. Xiong \textit{et al}}.
\newblock Towards a fast centralized controller for integrated silicon photonic
  multistage mzi-based switches.
\newblock In {\em OFC}, 2016.

\bibitem{opt_mitigation3}
{P. Guo \textit{et al}}.
\newblock Potential threats and possible countermeasures for photonic
  network-on-chip.
\newblock {\em IEEE Communications Magazine}, 2020.

\bibitem{sergio}
{C. Moratelli \textit{et al}}.
\newblock {The Convergence of Technologies to Provide Security on IoT Edge
  Devices}.
\newblock {\em Convergence}, 2021.

\bibitem{clements}
{W.R. Clements \textit{et al}}.
\newblock Optimal design for universal multiport interferometers.
\newblock {\em Optica}, 2016.

\bibitem{Kyatam:20}
{S. Kyatam \textit{et al}}.
\newblock Estimation of maximum temperature and thermal crosstalk between two
  active elements in a pic: development of a thermal equivalent circuit.
\newblock {\em Appl. Opt.}, 2020.

\bibitem{opt_mitigation}
{J. Zhou \textit{et al}}.
\newblock Attack mitigation of hardware trojans for thermal sensing via
  micro-ring resonator in optical nocs.
\newblock {\em JETC}, 2021.

\bibitem{lukas}
{L. Chrostowski and M. Hochberg}.
\newblock {\em {Silicon Photonics Design: From Devices to Systems}}.
\newblock Cambridge University Press, 2015.

\bibitem{dl_siph}
{F.P. Sunny \textit{et al}}.
\newblock A survey on silicon photonics for deep learning.
\newblock {\em J. Emerg. Technol. Comput. Syst.}, jun 2021.

\bibitem{mit_acc}
{Y. Shen \textit{et al}}.
\newblock {Deep learning with coherent nanophotonic circuits}.
\newblock {\em {Nature Photon 11}}, 2017.

\bibitem{9474000}
{S. Banerjee \textit{et al}}.
\newblock Modeling silicon-photonic neural networks under uncertainties.
\newblock In {\em DATE}, 2021.

\bibitem{4447311}
{S. R. Sarangi \textit{et al}}.
\newblock Varius: A model of process variation and resulting timing errors for
  microarchitects.
\newblock {\em IEEE TSM Journal}, 2008.

\bibitem{countering}
{Y. Zhu \textit{et al}}.
\newblock Countering variations and thermal effects for accurate optical neural
  networks.
\newblock In {\em IEEE/ACM ICCAD}, 2020.

\bibitem{puf}
{N. N. Anandakumar \textit{et al}}.
\newblock {Compact Implementations of FPGA-based PUFs with Enhanced
  Performance}.
\newblock In {\em {VLSID}}, 2017.

\bibitem{9465434}
{F. Pavanello \textit{et al}}.
\newblock Recent advances in photonic physical unclonable functions.
\newblock In {\em 2021 IEEE European Test Symposium (ETS)}, 2021.

\bibitem{25930692593144}
{D. M. Ancajas \textit{et al}}.
\newblock Fort-nocs: Mitigating the threat of a compromised noc.
\newblock In {\em Design Automation Conference}. ACM, 2014.

\bibitem{onn_under}
{S. Banerjee \textit{et al}}.
\newblock Characterizing coherent integrated photonic neural networks under
  imperfections.
\newblock {\em JLT}, 2022.

\bibitem{mahdi_pv}
{M. Nikdast \textit{et al}}.
\newblock {Chip-Scale Silicon Photonic Interconnects: A Formal Study on
  Fabrication Non-Uniformity}.
\newblock {\em {IEEE JLT Journal}}, 2016.

\bibitem{34478183460374}
{S. Yuan \textit{et al}}.
\newblock Pssm: Achieving secure memory for gpus with partitioned and sectored
  security metadata.
\newblock In {\em Proceedings of the ACM International Conference on
  Supercomputing}. ACM, 2021.

\bibitem{inter}
{INTERCONNECT from LUMERICAL Tools - Last access on 08/2022}.

\bibitem{benes}
V.E. Benes.
\newblock On rearrangeable threestage connecting networks.
\newblock Bell Syst., 1962.

\bibitem{spanke}
{R. A. Spanke and V. E. Benes}.
\newblock N-stage planar optical permutation network.
\newblock {\em Appl. Opt.}, 1987.

\bibitem{mcgill_acc}
{F. Shokraneh \textit{et al}}.
\newblock A single layer neural network implemented by a $4\times 4$ mzi-based
  optical processor.
\newblock {\em IEEE Photonics Journal}, 2019.

\bibitem{hyco}
{F. G. De Magalhaes \textit{et al}}.
\newblock {HyCo: A Low-Latency Hybrid Control Plane for Optical Interconnection
  Networks}.
\newblock In {\em RSP}, 2021.

\bibitem{hfrisc}
{F.T. Bortolon \textit{et al}}.
\newblock {Design and analysis of the HF-RISC processor targeting voltage
  scaling applications}.
\newblock In {\em {ACM SBCCI}}, 2016.

\end{thebibliography}

\end{document}